\documentclass[12pt]{article}
\pdfoutput=1
\usepackage{amsbsy}
\usepackage{amsfonts}
\usepackage{amsmath,amssymb}
\usepackage{bbold}
\usepackage{pdfpages}
\usepackage{comment}

\newcommand{\sect}[1]{\setcounter{equation}{0}\section{#1}}
\newcommand{\eq}{\begin{equation}}
\newcommand{\eqa}{\begin{eqnarray}}  
\newcommand{\en}{\end{equation}}
\newcommand{\ena}{\end{eqnarray}}
\newcommand{\enn}{\nonumber \end{equation}}

\def\sk{\vskip .4cm}
\def\noi{\noindent}

\def\al{\alpha}

\def\ga{\gamma}

\let \part\partial

\def\unmezzo{{1 \over 2}}
\def\epsi{\varepsilon}
\def\we{\wedge}

\def\de{\delta}

\def\part{\partial}

\def\sk{\vskip .4cm}

\def\noi{\noindent}

\def\X0{X^0}

\def\al{\alpha}
\def\ga{\gamma}

\def\unmezzo{{1 \over 2}}
\def\epsi{\varepsilon}
\def\epsibold{{\bf \epsilon}}

\def\we{\wedge}

\def\de{\delta}

\def\Dcal{{\cal D}}
\def\Rcal{{\cal R}}

\def\square{{\,\lower0.9pt\vbox{\hrule \hbox{\vrule height 0.2 cm
\hskip 0.2 cm \vrule height 0.2 cm}\hrule}\,}}

\def\epsilonbar{{\bar \epsilon}}
\def\thetabar{{\bar \theta}}

\def\Rboldtilde{\widetilde \Rbold}
\def\Omboldtilde{\widetilde \Ombold}
\def\Atilde{\widetilde A}
\def\epsiboldtilde{\widetilde \epsibold}
\def\Rtilde{\widetilde{R}}

\def\etatilde{{\widetilde \eta}}
\def\psibar{\bar \psi}
\def\epsilonbar{\bar \epsilon}

\def\Om{\Omega}

\def\Sigmabar{\overline \Sigma}

\def\Rbold{{\bf R}}

\def\Ombold{{\bf \Om}}
\def\onebold{{\bf 1}}
\def\epsibold{\boldsymbol {\epsi}}

\begin{document}

\begin{titlepage}
\rightline{ARC-2020-09}
%\rightline{hep-th/9509031}
%\rightline{November 2011} 
\vskip 2em
\begin{center}
{\LARGE \bf Chern-Simons Supergravity \\ \vskip 0.3cm on Supergroup Manifolds} \\[3em]

\vskip 0.5cm

{\bf
L. Castellani$^{1,2,3}$,  C. A.  Cremonini$^{4,5}$, and P. A. Grassi}$^{1,2,3}$
\bigskip

{\sl $^1$ Dipartimento di Scienze e Innovazione Tecnologica\\Universit\`a del Piemonte Orientale, viale T. Michel 11, 15121 Alessandria, Italy\\ [.5em] 
$^2$ INFN, Sezione di Torino, via P. Giuria 1, 10125 Torino, Italy\\ [.5em]
$^3$ Arnold-Regge Center, via P. Giuria 1, 10125 Torino, Italy\\ [.5em] 
$^4$ Dipartimento di Scienze e Alta Tecnologia (DiSAT), \\
Universit\`a degli Studi dell'Insubria, via Valleggio 11, 22100 Como, Italy\\ [.5em] 
$^5$ INFN, Sezione di Milano, via G.~Celoria 16, 20133 Milano, Italy
\\[4em]}
\end{center}

\begin{abstract}
\sk

We construct N=1 d=3 AdS supergravity within the group manifold approach and compare it with Achucarro-Townsend Chern-Simons formulation of the same theory. We clarify the relation between the off-shell super gauge transformations of the Chern-Simons theory and the off-shell worldvolume supersymmetry transformations of the group manifold action. We formulate the Achucarro-Townsend model 
in a double supersymmetric action where the Chern-Simons theory with a supergroup gauge symmetry 
is constructed on a supergroup manifold. This framework is useful to establish a correspondence of degrees of freedom and auxiliary fields between the two descriptions of  d=3 supergravity.

\end{abstract}

\vskip 3cm
 \noi \hrule \vskip .2cm \noi {\small
leonardo.castellani@uniupo.it \\ carlo.alberto.cremonini@gmail.com \\pietro.grassi@uniupo.it}

\end{titlepage}

\newpage
\setcounter{page}{1}

\tableofcontents

\vfill\eject

\sect{Introduction}

We consider the $N=1$ anti-de Sitter supergravity action in $d=3$, realized as the difference of two Chern-Simons actions \cite{Achucarro:1987vz}, with respectively $OSp(1|2)$ and $Sp(2)$ connections. Starting from the Chern-Simons formulation, we derive the supergravity action following the steps of the
Achucarro and Townsend construction. One obtains a theory whose fundamental 1-form fields are (after a simple redefinition) 
 the dreibein $V^a$, the spin connection $\omega^{ab}$ and the Majorana gravitino $\psi$. The action is invariant by construction under the gauge transformations of $OSp(1|2) \otimes Sp(2)$. The transformations generated by the spinorial (Majorana) charge of the supergroup yield the $N=1$ supersymmetry transformations, and close off-shell without need of auxiliary fields since they are really part of a gauge algebra. The action, being the integral of a 3-form on a 3-dimensional manifold, is also invariant by construction under 3d diffeomorphisms. The latter are distinct from the gauge symmetries generated by the translation charges of the gauge supergroup.
\sk
Next we consider the (super)group-geometric construction of $N=1$, $d=3$ anti-de Sitter supergravity \cite{if3}. In this framework the basic 1-form fields live on the whole supergroup manifold $OSp(1|2) \otimes Sp(2)$, and the spacetime coordinates are identified with 
the parameters of the translation subgroup. Here supersymmetry is realized as a diffeomorphism in this supergroup manifold, in the
fermionic directions. We will call it \emph{worldvolume supersymmetry} to distinguish it from the \emph{gauge supersymmetry} of the Chern-Simons action (see \cite{Zanelli}). To obtain a spacetime action (involving fields that depend only on spacetime coordinates), so as to be able
to compare it with the Achucarro-Townsend action, it is necessary to integrate out the dependence on the Lorentz and Grassmann coordinates
of the supermanifold. The resulting spacetime action coincides with the Achucarro and Townsend action, and is worldvolume supersymmetric provided some conditions are fulfilled, called
``rheonomic" conditions. We show how these conditions can be imposed as constraints on the ``outer" (i.e. along Grassmann differentials) components of the 2-form curvatures, and how this leads to a local supersymmetry  that not surprisingly
coincides with the gauge supersymmetry. Here the origin of supersymmetry is geometric, whereas
the gauge supersymmetry of the Chern-Simons action is totally algebraic. A peculiarity of the $d=3$ theory is that there
exist additional symmetries, due to other possible rheonomic conditions, that close only on-shell. In particular 
the action is also invariant under a supersymmetry that has an extra term in the spin connection transformation with respect to the gauge supersymmetry of the CS action.

Supergravity in $d=3$ AdS spacetime can also be formulated with an additional bosonic auxiliary field, to balance off-shell degrees of freedom (the superspace formulation can be found in \cite{Ruiz,Marcus,Gates}). In this formulation we find that worldvolume supersymmetry does not require all
curvatures to be horizontal in the Grassmann directions, and that Bianchi identities are satisfied off-shell.
The resulting theory extends the Achucarro-Townsend action with terms depending on the auxiliary field. Once the auxiliary field is eliminated via its (algebraic) field equation, the Achucarro-Townsend action is recovered.

%%%%%%%%%%%%%%%%%%%%%%%%%%%%

\sect{$N=1,d=3$ AdS supergravity as Chern-Simons }

%%%%%%%%%%%%%%%%%%%%%%%%%%%%

Here we treat the simplest $N=1$ case. We consider therefore the difference between a CS action for $OSp(1|2)$ and a CS action for $Sp(2)$:
\eq
S = \kappa \int_{M^3} STr( \Rbold \Ombold + {1 \over 3} \Ombold^3)- \kappa \int_{M^3} Tr ( \Rboldtilde \Omboldtilde + {1 \over 3} \Omboldtilde^3) \label{action1}
\en
where the 1-form $OSp(1|2)$ and $Sp(2)$ connections are given respectively by the $3 \times 3$ supermatrix  
$\Ombold$ and the $2 \times 2$ matrix $\Omboldtilde$:
 \eq
  \Ombold =
\left(
\begin{array}{cc}
  A^a \gamma_a & { 1 \over  \sqrt{\lambda}} \psi \\
 {i \over  \sqrt{\lambda}}  \psibar &  0  \\
\end{array}
\right), ~~~~\Omboldtilde = \Atilde^a \gamma_a 
    \label{Omdef}  
  \en
  and the 2-form curvatures are 
 \eq
      \Rbold =  d \Ombold - \Ombold \we \Ombold~
  \equiv  \left(
\begin{array}{cc}
   R^a (A) \gamma_a  & {1 \over  \sqrt{\lambda}} \Sigma\\
 {i \over  \sqrt{\lambda}}  \Sigmabar  &  0 \\
\end{array}
\right) ,~~~~\Rboldtilde =  \Rtilde^a(\Atilde) \gamma_a \label{Rdef}
       \en
\noi with
   \eqa
    & & R^a(A) = dA^a + \epsi^a_{~bc} A^b A^c - {1 \over 2 \lambda}  \psibar \gamma^a \psi  \label{defR}\\
    & & \Rtilde^a(\Atilde) = d \Atilde^a + \epsi^a_{~bc} \Atilde^b \Atilde ^c   \label{defRtilde}\\
    & & \Sigma = d \psi - A^a \gamma_a \psi  \label{defSigma} \\
    & & \Sigmabar = d \psibar - \psibar \gamma_b A^b  
    \ena
    Carrying out the traces in (\ref{action1}) leads to the action
    \eq
    S = 2 \kappa \int_{M^3} R^a(A) A_a - \Rtilde^a (\Atilde) \Atilde_a - {1 \over 3} (A^a A^b A^c - \Atilde^a \Atilde^b \Atilde^c ) \epsi_{abc} - \psibar \Sigma \label{action1bis}
    \en
Defining now the dreibein $V^a$ and the spin connection $\omega^{ab}$ as combinations of the 
$A^a$ and $\Atilde^a$ connections:
\eq
A^a = \unmezzo ( \omega^a + {1 \over  \lambda} V^a),~~~~ \Atilde^a = \unmezzo (\omega^a - {1 \over  \lambda} V^a)
\label{Acombinations} \en
 the action (\ref{action1bis}) becomes:
\eq
S = - {\kappa \over \lambda} \int_{M^3} (\Rcal^{ab} V^c \epsi_{abc} - {1 \over 3 \lambda^2} V^a V^b V^c +  2 i \psibar \Sigma )
\label{action2}
\en
where 
\eq
\Rcal^{ab} \equiv d \omega^{ab} - \omega^a_{~c} \omega^{cb}
\en
is the Lorentz curvature, and the gravitino curvature $\Sigma$ is expressed as
\eq
\Sigma = d \psi - {1 \over 4} \omega^{ab} \gamma_{ab} \psi - {1 \over 2 \lambda} V^a \gamma_a \psi
\en
\noi {\bf Symmetries}
\sk
The action (\ref{action1}) or equivalently (\ref{action2}) is invariant (up to boundary terms) under the
gauge transformations:
\eqa
& \delta_{\epsibold}  \Ombold = d \epsibold - \Ombold  \epsibold + \epsibold \Ombold,~~~ \Rightarrow ~~~\de_{\epsibold} \Rbold = - \Rbold \epsibold + \epsibold \Rbold  \label{gaugetransf1} \\
& \delta_{\epsiboldtilde}  \Omboldtilde = d \epsiboldtilde - \Omboldtilde  \epsiboldtilde + \epsiboldtilde \Omboldtilde,~~~ \Rightarrow ~~~\de_{\epsiboldtilde} \Rboldtilde = - \Rboldtilde \epsiboldtilde + \epsiboldtilde \Rboldtilde  \label{gaugetransf2} 
 \ena
where $\epsibold$ and $\epsiboldtilde$ are the $OSp(1|2)$ and $Sp(2)$ gauge parameters:
 \eq
      \epsibold =   \left(
\begin{array}{cc}
   \eta^a  \gamma_a  & {1 \over \sqrt{\lambda}} \epsilon\\
{i \over \sqrt{\lambda}} \epsilonbar  &  0 \\
\end{array}
\right) ,~~~~\epsiboldtilde =  \etatilde^a \gamma_a \label{Rdef2}
       \en
       On the component fields the gauge transformations (\ref{gaugetransf1}) and (\ref{gaugetransf2}) take the form:
       \eqa
      & & \delta A^a = d \eta^a + 2 A^b \eta^c \epsi^a_{~bc} + {i \over \lambda} \epsilonbar \gamma^a \psi \\
        & & \delta \Atilde^a = d \etatilde^a + 2 A^b \etatilde^c \epsi^a_{~bc} \\
        & & \delta \psi = d \epsilon - A^a \gamma_a \epsilon + \eta^a \gamma_a \psi
         \ena
Using now the definitions (\ref{Acombinations}), the gauge transformations on the supergravity fields
read:
\eqa
& & \delta V^a = \Dcal \epsi^a + \epsi^a_{~b} V^b + i \epsilonbar \gamma^a \psi \label{gaugeVa}\\
& & \delta \omega^{ab} = \Dcal \epsi^{ab} - {2 \over \lambda^2} V^{[a} \epsi^{b]} - {i \over \lambda} \epsilonbar \gamma^{ab} \psi \\
& & \delta \psi = \Dcal \epsilon - {1 \over 2 \lambda} V^a \gamma_a \epsilon + {1 \over 4} \epsi^{ab} \gamma_{ab} \psi + {1 \over 2 \lambda} \epsi^a \gamma_a \psi \label{gaugepsi}
\ena
The translation and Lorentz rotation parameters $\epsi^a$ and $\epsi^{ab}$ are defined in terms of $\eta^a$ and $\etatilde^a$ as
\eq
\epsi^a \equiv  \lambda (\eta^a - \etatilde^a),~~~\epsi^{ab} \equiv \epsi^{ab}_{~~c} (\eta^c + \etatilde^c)
\en
and $\Dcal$ is the Lorentz covariant derivative:
\eqa
& & \Dcal \epsi^a \equiv d \epsi^a - \omega^a_{~b} V^b \\
& & \Dcal \epsi^{ab} \equiv d \epsi^{ab} - \omega^a_{~c} \epsi^{cb} + \omega^b_{~c} \epsi^{ca} \\
& & \Dcal \epsilon \equiv d \epsilon- {1 \over 4} \omega^{ab} \gamma_{ab} \epsilon
\ena

\sk
\noi {\bf Field equations}
\sk
Varying the action (\ref{action2}) in  $V^a$, $\omega^{ab}$ and $\psi$ leads to the field equations:
\eqa
& & \Rcal^{ab} - {1 \over \lambda^2} V^a V^b + {i \over 2 \lambda} \psibar \gamma^{ab} \psi =0 \\
& & \Dcal V^a - {i \over 2} \psibar \gamma^a \psi =0 \\
& &  \Dcal \psi - {1 \over 2 \lambda} V^a \gamma_a \psi =0
\ena
The left hand sides are the curvatures of the $OSp(1|2) \times Sp(2)$ supergroup, in the rotated basis
($V^a,\omega^{ab}, \psi$). These equations in fact are just the Cartan-Maurer equations of the supergroup,
and are the starting point of the group-geometric construction of next Section.

%%%%%%%%%%%%%%%%%%%%%%%%%%%%%%%%%%%

\sect{$N=1,d=3$ AdS supergravity in the group geometric approach}

%%%%%%%%%%%%%%%%%%%%%%%%%%%%%%%%%%%

The $OSp(1|2) \times Sp(2)$ Cartan-Maurer equations yield the definitions of the super-AdS curvatures:
   \eqa\label{d3cm1}
  & & R^{ab}=d \omega^{ab} - \omega^a_{~c} ~ \omega^{cb} - {1 \over \lambda^2} V^a V^b + {i \over 2 \lambda} \psibar \gamma^{ab} \psi\\
   & & R^a=dV^a - \omega^a_{~b} ~ V^b - {i \over 2} \psibar \gamma^a \psi \equiv \Dcal V^a - {i \over 2} \psibar \gamma^a \psi\ \label{torsionRa}\\
   & & \Sigma = d \psi - {1 \over 4} \omega^{ab} \gamma_{ab} \psi - {1 \over 2 \lambda} V^a \gamma_a \psi = 
   \Dcal \psi - {1 \over 2 \lambda} V^a \gamma_a \psi \label{d3cm3}
    \ena
The Cartan-Maurer equations are invariant under the rescalings
     \eq
     \omega^{ab} \rightarrow  \omega^{ab}, ~V^a \rightarrow u V^a,~\psi \rightarrow u^{1\over 2} \psi,~\lambda \rightarrow u \lambda   \label{rescalings3}
           \en
     Taking exterior derivatives of both sides yields the Bianchi identities:
     \eqa
    & &  \Dcal R^{ab} + {2 \over \lambda^2} R^{[a} V^{b]} + {i \over \lambda} \psibar \gamma^{ab} \Sigma =0 \label{bianchiRab1} \\
    & &  \Dcal R^a + R^a_{~b} ~ V^b - i~ \psibar \gamma^a \Sigma =0 \label{bianchiRa1}\\
    & & \Dcal \Sigma + {1 \over 4} R^{ab} \gamma_{ab} ~\psi + {1 \over 2 \lambda} R^a \gamma_a \psi - {1 \over 2 \lambda} V^a \gamma_a \Sigma=0   \label{bianchiSigma1}  \ena
    
     \subsection{The Lagrangian}

Applying the building rules of the geometric approach \cite{cube} yields the Lagrangian 3-form
          \eqa
     & &  L =  R^{ab} V^c \epsi_{abc} +  2 i \psibar \Sigma  + {2 \over 3 \lambda^2} V^a V^b V^c  - {i \over 2 \lambda} \psibar \gamma^{ab} \psi V^c \epsi_{abc}  = \nonumber \\
     & & ~~~~~=  \Rcal^{ab} V^c \epsi_{abc} - {1 \over 3 \lambda^2} V^a V^b V^c +  2 i \psibar \Sigma
           \label{d3lagrangian}
            \ena
 and is formally identical to the Achucarro-Townsend Lagrangian of the previous Section. Note however that in 
 the present framework this 3-form lives on the whole $N=1$ superspace $M^{3|2}$ . It is obtained by considering the most general Lorentz scalar 3-form, given in terms of the super AdS curvatures
            and fields, invariant under the rescalings discussed above, and such that
        the variational equations admit the vanishing curvatures solution
             \eq
             R^{ab} = R^a = \Sigma = 0
             \en

     \subsection{Action and symmetries}
     
     The action is now an integral over the whole superspace $M^{3|2}$
     \eq
     S = \int_{M^{3|2}} L \wedge \eta_{M^3} ~ \label{actionsuperspace}
     \en
     and $\eta_{M^3}$ is the Poincar\'e dual of the 3-dimensional Minkowski space $M^3$ immersed into the superspace $\mathcal{M}^{(3|2)}$ (see e.g. \cite{if1}). $\eta_{M^3}$ is a 2-form in superspace\footnote{$\eta_{M^3}$ is really a $(0|2)$-integral form in superspace and the Lagrangian has to be considered as a $(3|0)$-superform so that the whole integrand is a $(3|2)$ top form in superspace.}
     that after integration localizes the
     Lagrangian on the $d=3$ bosonic subspace, i.e.
     \eq
  S= \int_{M^{3}} L_{\theta = d \theta=0}  \label{actionspacetime}
     \en
     The action then exactly reproduces the Achucarro-Townsend spacetime action. Written as in (\ref{actionsuperspace}) 
     the action is automatically invariant under superdiffeomorphisms in superspace, since it is a 5-form integrated
     on a 5-dimensional superspace. The superdiffeomorphisms along a tangent vector $v$ in superspace act on the 1-form fields in the Lagrangian $L$ with the Lie  derivative  $\ell_v \equiv d \iota_v + \iota_v d$, i.e.
     \eqa
     & & \delta_v V^a = d (\iota_v V^a) + \iota_v dV^a = d (\iota_v V^a) + \iota_v R^a + \iota_v (\omega^a_{~b} V^b)  +{i \over 2}  \iota_v (\psibar \gamma_a \psi) \nonumber \\
     & & ~~~~~~ =  \Dcal (\iota_v V^a) +  \iota_v R^a + \iota_v (\omega^a_{~b}) V^b +{i \over 2}  \iota_v (\psibar \gamma_a \psi)   \label{diffV}
     \ena
     and similarly
     \eqa
     & &\delta_v \omega^{ab} =  \Dcal (\iota_v \omega^{ab}) +  \iota_v R^{ab}  
          + {1 \over \lambda^2} \iota_v (V^a V^b) - {i \over 2 \lambda} \iota_v (\psibar \gamma^{ab} \psi) \label{diffomega}  \\
          & &   \delta_v \psi =  \Dcal (\iota_v \psi) +  \iota_v \Sigma +{1 \over 4} \iota_v (\omega^{ab}) \gamma_{ab} \psi + {1 \over 2 \lambda} \iota_v (V^a \gamma_a \psi ) \label{diffpsi}
          \ena
   These are the built-in invariances of the action (\ref{actionsuperspace}). Here resides most of the power of the group manifold formalism: if one considers the ``mother"  action (\ref{actionsuperspace}) on $M^{3|2}$, the guaranteed symmetries are {\it all} the diffeomorphisms on $M^{3|2}$, generated by the Lie derivative $\ell_v$ along the tangent vectors  $v$ of $M^{3|2}$.
  But how do these symmetries transfer to the spacetime action (\ref{actionspacetime}) ?
  
  The variation of the superspace action under diffeomorphisms generated by $\ell_v$ is\footnote{Recall  that $\ell_v$(top form) = $d(\iota_v$ top form).}
  \eq
   \delta S = \int_{M^{3|2}}  \ell_v (L \wedge \eta_{M^3} )= \int_{M^{3|2}}  (\ell_v L) \wedge \eta_{M^3} + L \wedge \ell_v \eta_{M^3} =0
   \en
  modulo boundary terms. One has to vary the fields\footnote{Since $\ell_v$ satisfies the Leibniz rule,   $\ell_v L$ can be computed by varying in turn all fields inside $L$.} in $L$ as well as the submanifold embedded in $M^{3|2}$:  the sum of these two variations gives zero\footnote{In the following the vanishing of action variations will always be understood modulo boundary terms.} on the superspace action $S$.  
    But what we need
   in order to have a {\it spacetime} interpretation of all the symmetries of $S$, is really
   \eq
   \delta S = \int_{M^{3|2}} (\ell_v L) \wedge \eta_{M^3} =0 \label{spacetimesymm}
   \en
 If this holds, varying the fields $\phi$ inside $L$ with the Lie derivative $\ell_v$ as in (\ref{diffV})-(\ref{diffpsi}), and then projecting on spacetime
 ($\theta=0,d\theta=0$),  yields spacetime variations
 \eq
\de \phi  (\theta=0,d\theta=0)) =  \ell_v \phi (x,\theta) |_{\theta=0,d\theta=0)}
 \en
  that leave the spacetime action (\ref{actionspacetime}) invariant. We call them
 {\it spacetime invariances}. They originate from the diffeomorphism invariance of the group manifold action,
 and give rise to symmetries of the spacetime action  (\ref{actionspacetime}) only when (\ref{spacetimesymm})
 holds.  This happens when $\ell_v L$ is exact, since $\eta$ is closed \cite{if3}. Exactness of  $\ell_v L$ is equivalent 
 to the condition
 \eq
 \iota_v dL = d \alpha  \label{idL}
 \en
 The Lagrangian $L$ depends on the basic fields $V^a, \omega^{ab}, \psi$ and their AdS curvatures  $R^A = R^a, R^{ab}, \Sigma$    so that also $dL$, using the Bianchi identities, is expressed in terms of the fields and their curvatures. Then condition (\ref{idL})  translates
 into  {\it conditions on the contractions} $\iota_v R^A$, i.e.  conditions on the curvature components. In the
 jargon of the group-geometric approach, these are called ``rheonomic" conditions, and must be consistent with Bianchi identities.
The symmetry transformations of the theory are then given by equations (\ref{diffV})-(\ref{diffpsi}), where the contractions $\iota_v R^a, \iota_v R^{ab}, \iota_v \Sigma$
are replaced by their expressions given by the rheonomic conditions.

 \subsection{Curvature parametrizations and symmetries of the spacetime action}

Computing the exterior derivative of the Lagrangian in (\ref{d3lagrangian}) and using the Bianchi identities (\ref{bianchiRab1})-(\ref{bianchiSigma1}) yields:
\eq
dL = R^{ab} R^c \epsi_{abc} + 2i~\Sigmabar \Sigma
\en
The condition 
\eqa
i_\epsilon dL = d\alpha \label{idL2}
\ena
where $\epsilon$ is a tangent vector in fermionic directions is satisfied if all curvatures have no ``legs" in fermionic directions,
i.e. if $i_\epsilon R^A=0$.  This leads to the parametrizations of the curvatures
\eqa
 & & R^{ab} = R^{ab}_{~~cd} V^a V^b \label{paramRab1}\\
 & & R^a=R^a_{~bc} V^b V^c  \\
 & & \Sigma = \Sigma_{ab} V^a V^b \label{paramSigma1}
 \ena
 and the transformations generated by the Lie derivative along the supergroup directions are:
 \eqa
& & \delta V^a = \delta_{gauge} ~V^a + 2 \epsi^b R^a_{~bc} V^c  \label{LieVa}\\
& & \delta \omega^{ab} = \delta_{gauge}~ \omega^{ab}  + 2 \epsi^c R^{ab}_{~~cd} V^d \\
& & \delta \psi = \delta_{gauge} ~\psi + 2 \epsi^a \Sigma_{ab} V^b \label{Liepsi}
\ena 
where $\delta_{gauge}$ are the gauge variations of the Achucarro -Townsend action given in (\ref{gaugeVa})-(\ref{gaugepsi}).
The difference are terms proportional to the AdS curvatures: these terms are necessary for the transformation parametrized 
by $\epsi^a$ to be a spacetime diffeomorphism, rather than a gauge translation. 

The spacetime reduced action being equal to the Achucarro-Townsend action, it also has its gauge symmetries.
These coincide with the ones expressed by eq.s (\ref{LieVa})-(\ref{Liepsi}), except for gauge translations, that are
an additional symmetry. In other words, the action is invariant under the CS gauge symmetry  (\ref{gaugeVa})-(\ref{gaugepsi}),
and ordinary spacetime diffeomorphisms. 

 \sk 
 But there is an additional symmetry, due to another solution of (\ref{idL2}), provided by the parametrizations:
\eqa
         & & R^{ab} = R^{ab}_{~~cd} ~V^c V^d + \thetabar^{ab}_{~~c}~\psi V^c  \label{parRab2} \\
        & & R^a = 0  \label{parRa2}\\
        & & \Sigma = \Sigma_{ab} V^a V^b  \label{paramSigma2}
        \ena
        with
         \eqa
         \thetabar^{ab}_{~~c} = c_1 ~\Sigmabar_c^{~[a} \gamma^{b]}+ c_2~ \Sigmabar^{ab} \gamma_c
         \ena
         The coefficients $c_1,c_2$ are fixed by the Bianchi identity (\ref{bianchiRa1})
        to the values:
         \eq
          c_1 = 2i,~c_2=-i
         \en
         However the other Bianchi identities hold only on-shell, i.e. for $R^{ab}=R^a=\Sigma=0$. Then the invariances
         generated by the Lie derivative:
         \eqa
         & & \delta V^a = \delta_{gauge} ~V^a  \label{LieVa2}\\
& & \delta \omega^{ab} = \delta_{gauge}~ \omega^{ab}  + 2 \epsi^c R^{ab}_{~~cd} V^d + \thetabar^{ab}_{~~c} \epsilon V^c
  - \thetabar^{ab}_{~~c}  \psi \epsi^c\\
& & \delta \psi = \delta_{gauge} ~\psi + 2 \epsi^a \Sigma_{ab} V^b \label{Liepsi2}
\ena 
         are still invariances of the action, but only close on-shell.

 Notice that the origin of supersymmetry is completely algebraic for the Chern-Simons
action, while it is geometric (due to superdiffeomorphism invariance of a superspace action) for the rheonomic action.

%%%%%%%%%%%%%%%%%%%%%%

\sect{Off-shell $N=1,d=3$ AdS supergravity}

%%%%%%%%%%%%%%%%%%%%%%

\subsection{Off-shell degrees of freedom}

The  mismatch between the 3 off-shell bosonic degrees of freedom of the dreibein ($d(d-1)/2$ in $d$ dimensions), and the 4 off-shell degrees of freedom of the gravitino ($(d-1)2^{[d/2]}$ in $d$ dimensions for Majorana or Weyl) can be cured by introducing an extra bosonic d.o.f., here provided by a bosonic 2-form auxiliary field $B$.

\subsection{The extended superAdS algebra}

The algebraic starting point is the FDA (Free Differential Algebra, see \cite{cube}) that enlarges the $d=3$ superAdS Cartan-Maurer equations
to include the auxiliary 2-form field $B$. This extension of the superAdS algebra is
possible due to the existence of the $d=3$ cohomology class $\Omega= \psibar \gamma_a \psi V^a$,
which is closed because of the $d=3$  Fierz identity (\ref{Fierz3d}).

The FDA yields the definitions of the AdS Lorentz curvature, the supertorsion, the AdS gravitino field strength and the 2-form field strength:
   \eqa
  & & R^{ab}=d \omega^{ab} - \omega^a_{~c} ~ \omega^{cb} - {1 \over \lambda^2} V^a V^b + {i \over 2 \lambda} \psibar \gamma^{ab} \psi \\
   \label{TORSION}& & R^a=dV^a - \omega^a_{~b} ~ V^b - {i \over 2} \psibar \gamma^a \psi \equiv \Dcal V^a - {i \over 2} \psibar \gamma^a \psi\ \\
   & & \Sigma = d\psi - {1 \over 4} \omega^{ab} \gamma_{ab} - {1 \over 2 \lambda} V^a \gamma_a \psi ~ \psi \equiv  \Dcal \psi   - {1 \over 2 \lambda} V^a \gamma_a \psi \\
   & & R^{\otimes}=dB-{i \over 2} \psibar \gamma^a \psi ~V^a  + {1 \over 3 \lambda} V^a V^b V^c \epsi_{abc} 
    \ena
The generalized Cartan-Maurer equations are invariant under the rescalings
     \eq
     \omega^{ab} \rightarrow  \omega^{ab}, ~V^a \rightarrow u  V^a,~\psi \rightarrow u^{1\over 2} \psi,~B \rightarrow u^2 B \label{rescalings4}
           \en
     Taking exterior derivatives of both sides yields the Bianchi identities:
     \eqa
    & &  \Dcal R^{ab} + {2 \over \lambda^2} R^{[a} V^{b]} + {i \over \lambda} \psibar \gamma^{ab} \Sigma =0 \label{bianchiRab2} \\
    & &  \Dcal R^a + R^a_{~b} ~ V^b - i~ \psibar \gamma^a \Sigma =0 \label{bianchiRa2}\\
    & & \Dcal \Sigma + {1 \over 4} R^{ab} \gamma_{ab} ~\psi + {1 \over 2 \lambda} R^a \gamma_a \psi - {1 \over 2 \lambda} V^a \gamma_a \Sigma=0   \label{bianchiSigma2}  \\
    & & dR^{\otimes} - i~ \psibar \gamma^a \Sigma V^a + {i \over 2} \psibar \gamma^a \psi~ R^a - {1 \over \lambda} R^a V^b V^c \epsi_{abc} = 0 \label{BianchiH2}
     \ena
    
      \subsection{Curvature parametrizations}

As in the preceding Section, we impose some algebraic constraints on the curvature components to ensure invariance
of the spacetime action under local supersymmetry. In this case, the presence of the auxiliary field $B$ allows
the off-shell closure of the symmetry algebra and now Bianchi identities hold also off-shell. The required
parametrization is given by
       \eqa
         & & R^{ab} = R^{ab}_{~~cd} ~V^c V^d + \thetabar^{ab}_{~~c}~\psi ~V^c + c_1~ f ~\psibar \gamma^{ab} 
         \psi \label{parRab3} \\
        & & R^a = 0  \label{parRa3}\\
        & & \Sigma = \Sigma_{ab} V^a V^b - c_2~f~V^a \gamma_a \psi  \\
        & & R^{\otimes} = f~V^a V^b V^c \epsilon_{abc} \\
        & & df = \partial_a f~ V^a + \psibar \Xi \label{pardf2}
        \ena
        with
         \eqa
         \thetabar^{ab}_{~~c} = c_3 ~\Sigmabar_c^{~[a} \gamma^{b]} + c_4 \Sigmabar^{ab} \gamma_c ~,~~~~~~~~~
         \Xi^\al =c_5 ~ \epsilon^{abc} \gamma_a \Sigma_{bc}
         \ena
         The coefficients $c_1,c_2, c_3, c_4,c_5$ are fixed by the Bianchi identities
        to the values:
         \eq
         c_1=  {3i \over 2} ,~c_2= {3 \over 2} ,~c_3 = 2i,~c_4=-i,~c_5 = - {i \over 3!}
         \en
         The $VVV$ component $f$ of $R^{\otimes}$ scales as $f \rightarrow u^{-1} f$,
         and is identified with the auxiliary scalar superfield of the superspace approach of ref. \cite{Ruiz}.
         Thanks to the presence of the auxiliary field, the Bianchi identities do not imply
         equations of motion for the spacetime components of the curvatures.
         
         \subsection{The Lagrangian}

     The usual building rules of the geometric approach lead to the Lagrangian 3-form
         \eqa
     & &  L =  R^{ab} V^c \epsi_{abc} +  2 i \psibar \Sigma  + {2 \over 3 \lambda^2} V^a V^b V^c  - {i \over 2 \lambda} \psibar \gamma^{ab} \psi V^c \epsi_{abc}  + \nonumber \\
     & & ~~~~~+ \alpha (f R^{\otimes} - {1 \over 2} f^2 V^a V^b V^c \epsi_{abc} ) \label{d3lagrangianaux}
            \ena
           The remaining parameter is fixed to $\alpha = 6$ by requiring $\iota_\epsilon dL^{3|0} = exact$, i.e. 
           supersymmetry invariance of
           the spacetime action. Indeed with $\alpha=6$ 
           we find $dL=0$
           on the (off-shell) field configurations satisfying the curvature parametrizations (\ref{parRab3})-(\ref{pardf2}).
           
     \subsection{Off-shell supersymmetry transformations}
       
           The off-shell closure of the supersymmetry transformations
           is ensured because the Bianchi identities hold without recourse to the 
           spacetime field equations. The action is invariant under these transformations, given by the Lie derivative of the fields along  the fermionic directions:
                \eqa
              & &  \delta_\epsilon V^a = -i \psibar \gamma^a \epsilon\\
              & &  \delta_\epsilon  \psi = \Dcal \epsilon - {1\over \ 2 \lambda}  V^a \gamma_a \epsilon + {3 \over 2} f  V^a \gamma_a \epsilon\\
              & & \delta_\epsilon \omega^{ab}= \thetabar^{ab}_{~~c} ~ \epsilon V^c - 3i f~ \psibar \gamma^{ab} \epsilon + {i \over \lambda} \psibar \gamma^{ab} \epsilon \\
              & & \delta_\epsilon  B = - i \psibar \gamma^a \epsilon V^a \\
              & & \delta_\epsilon f = \epsilonbar ~ \Xi
                \ena

     \subsection{Field equations}
  
   Varying $\omega^{ab}$, $V^a$, $\psi$, $B$ and $f$ in the action $\displaystyle S = \int_{M^{3|2}} L^{3} \wedge \eta_{M^3} $ leads to the equations of motion:
                \eqa \label{reoEQ}
                 & & R^a=0 \\
                 & & R^{ab} = 9 f^2 V^a V^b - {6 f \over \lambda} V^a V^b + {3i \over 2} f ~ \psibar \gamma^{ab} \psi \\
                 & & \Sigma = - {3 \over 2} V^a  \gamma_a \psi f \\
                 & & df=0 \\
                 & & R^{\otimes}= f~V^a V^b V^c \epsi_{abc}
                 \ena

%%%%%%%%%%%%%%%%%%%%%%%%%%%%%%%%%%%%%
 
\sect{Equivalence of transformations: trivial gauge transformations.}

%%%%%%%%%%%%%%%%%%%%%%%%%%%%%%%%%%%%%

Here we show that the gauge transformations of super CS action are equivalent to the diffeomorphism transformations on the supergravity counterpart, modulo \emph{trivial gauge transformations}, i.e. transformations which are proportional to the equations of motion.

Let us start from the CS side: we analyse first the bosonic symmetries and therefore we restric ourselves to a pure gravity theory. From the gauge fields $A$ and $\tilde{A}$ we can obtain the dreibein $V^a$ 
and the spin connection $\omega^{ab}$ as linear combinations of $A$ and $\tilde{A}$. Let us focus on their bosonic gauge transformations:
\begin{equation}\label{ETA}
	\delta V^a = \mathcal{D} \varepsilon^a + \varepsilon^{ab} V_b \ , \ 
	\delta \omega^{ab} = \mathcal{D} \varepsilon^{ab}
\end{equation}
On the other hand, let us see how $V$ transforms under diffeomorphisms and Lorentz symmetries:
\begin{equation}\label{ETB}
	\tilde{\delta} V^a = \ell_X V^a + \lambda^a_{~b} V^b \ , \ 
	\tilde{\delta} \omega^{ab} = \ell_X \omega^{ab} + \mathcal{D} \lambda^{ab} 
\end{equation}
where $X$ is a vector field and $\lambda^{a}_b$ are the Lorentz parameters. We can recast the transformations in \eqref{ETB} as follows:
\begin{align}
	\nonumber \tilde{\delta} V^a & = \ell_X V^a + \omega^a_{~b} \wedge V^b \nonumber \\ &=
	\iota_X \left( d V^a - \omega^a_{~b} \wedge V^b \right) + \iota_X \left( \omega^a_{~b} \wedge V^b \right) + \mathcal{D} \left( \iota_X V^a \right) + \omega^a_{~b} \wedge \iota_X V^b + \lambda^a_{~b} \wedge V^b  
	\nonumber \\
	\label{ETC} & = \iota_X R^a + \iota_X \omega^a_{~b} \wedge V^b + 
	\mathcal{D} \left( \iota_X V^a \right) + \lambda^a_{~b} \wedge V^b \nonumber \\ 
	&= \iota_X R^a + 
	\mathcal{D} \left( \iota_X V^a \right) + \left( \lambda + \iota_X \omega \right)^a_{~b} \wedge V^b \ .
\end{align}
We have therefore distinguished three pieces: the first one is written in terms of the \emph{torsion} $R^a$, the second may be identified with $\mathcal{D} \varepsilon^a$ if we identify
\begin{equation}\label{ETD}
	\iota_X V^a \equiv \varepsilon^a \ \implies \ X^\mu V_\mu^a \equiv \varepsilon^a \ ,
\end{equation}
and the last term is identified with $\varepsilon$ if we identify
\begin{equation}\label{ETE}
	\varepsilon \equiv \lambda + \iota_X \omega \ .
\end{equation}
Therefore, the difference between the gauge and Lorentz transformations and the diffeomorphisms and 
Lorentz transformations is
\begin{equation}\label{ETF}
	\left( \delta - \tilde{\delta} \right) V^a = - \iota_X R^a \ .
\end{equation}
We can perform the same manipulations of \eqref{ETC} for the spin connection as well:
\begin{equation}\label{ETG}
	\tilde{\delta} \omega^{ab} = \iota_X \left( d \omega - \omega \wedge \omega \right)^{ab} + 
	\mathcal{D} \left( \iota_X \omega^{ab} \right) + \mathcal{D} \Omega^{ab} = 
	\iota_X R^{ab} + \mathcal{D} \left(\lambda + \iota_X \omega \right)^{ab} \ .
\end{equation}
The first term is written in terms of the \emph{curvature} $R^{ab}$, 
while the second term can be identified with \eqref{ETB} by setting
\begin{equation}\label{ETH}
	\varepsilon \equiv \lambda + \iota_X \omega \ .
\end{equation}
Again, the difference between gauge + Lorentz and diffeomorphism + Lorentz transformations reads
\begin{equation}\label{ETI}
	\left( \delta - \tilde{\delta} \right) \omega^{ab} = - \iota_X R^{ab} \ .
\end{equation}
Recall that, given the Einstein-Hilbert action, we can recast $R^{ab}$ and $R^a$ as
\begin{equation}\label{ETJ}
	\frac{\delta S}{\delta V^a} = \epsilon_{a b c} R^{b c} \ , \ \frac{\delta S}{\delta \omega^{a b}} = 
	\epsilon_{a b c} R^c \ ,
\end{equation}
We can therefore recast \eqref{ETF} and \eqref{ETI} as
\begin{equation}
	\label{ETK} \left( \delta - \tilde{\delta} \right) V^a_\mu  = \epsilon^{a b c} X_\nu \epsilon_\mu^{~\nu \rho} \frac{\delta S}{\delta \omega_\rho^{bc}} \,, ~~~~~
	 \left( \delta - \tilde{\delta} \right) \omega^{a b}_\mu  = \epsilon^{a b c} X_\nu \epsilon_\mu^{~\nu \rho} \frac{\delta S}{\delta V_\rho^c} \ .
\end{equation}
These kind of transformations, defined by a parameter multiplying the equations of motions, are called \emph{trivial gauge transformations}; any action is invariant under these transformations
and they can be cast in the following 
from 
\begin{eqnarray}
\label{ETKA}
\delta \phi^A = \mu^{AB} \frac{\delta S}{\delta \phi^B}
\end{eqnarray}
for any field $\phi^A$ of the model. The gauge parameters $\mu^{AB}$ are local, possibly field-dependent 
gauge parameters. They are antisymmetric $\mu^{AB} = - \mu^{BA}$ and leave any action invariant 
\begin{eqnarray}
\label{ETKB}
\delta S = \int \delta\phi^A  \frac{\delta S}{\delta \phi^A} = \int  
\mu^{AB} \frac{\delta S}{\delta \phi^B} \frac{\delta S}{\delta \phi^A} = 0 \,. 
\end{eqnarray}
The commutator of any gauge transformation of the theory
\begin{eqnarray}
\label{ETKC}
\delta_T S = T^A \frac{\delta S}{\delta \phi^A} =0\,,
\end{eqnarray}
with trivial gauge transformations (\ref{ETKA})
\begin{eqnarray}
\label{ETKD}
[\delta_\mu, \delta_T] \phi^A = 
\left( \frac{\delta T^A}{\delta \phi^B} \mu^{BC}  - \frac{\delta T^C}{\delta \phi^B} \mu^{BA}  
- T^A  \frac{\delta \mu^{BC}}{\delta \phi^B} \right)  \frac{\delta S}{\delta \phi^C} 
  \end{eqnarray}
leads again to a trivial gauge transformation. 
This set of trivial gauge transformations forms a normal (i.e. invariant) subgroup of the full gauge group. 
They are not physically relevant since: 1) 
they exist independently of the action; in other words, they do not restrict at all the form of the Lagrangian and no non-trivial Noether identity is associated with them. 2) they imply no degeneracy of the action and in the Hamiltonian formalism, there is no corresponding constraint. Actually, the conserved charges associated with those gauge transformation, when rewritten as phase space functions, vanish identically. 3) The trivial gauge 
transformations vanish on-shell, i.e., do not map solutions of the equations of motion on new, different solutions. 4) There is accordingly no need for a "gauge fixing". On this basis, it is natural to disregard them 
and, being a normal subgroup, this is well defined procedure \cite{Henneaux:1992ig,Witten,Achucarro2}.

Notice that \eqref{ETF} and \eqref{ETH}, once inserted into the action, give rise to
\begin{equation}
	\iota_{X} \left( R^a R^{b c} \epsilon_{a b c} \right) = \iota_X \left( d L \right) = 0 \ ,
\end{equation}
since $d L =0$ being $L$ a 3-form in a 3-dimensional manifold.

Now, we proceed with the local supersymmetry transformations for $d=3$ supergravity. We write the susy transformations as Lie derivatives along fermionic directions:
\begin{equation}\label{ETL}
	\tilde{\delta} V^a = \ell_\epsilon V^a = d \iota_\epsilon V^a + \iota_\epsilon d V^a = \iota_\epsilon R^a + \iota_\epsilon \left( \omega \wedge V \right)^a + \iota_\epsilon \left( \frac{i}{2} \bar{\psi} \gamma^a \psi \right) = $$ $$ = \iota_\epsilon R^a + \left[ \iota_\epsilon \left( \omega \right) V \right]^a + i \epsilon \gamma^a \psi = \iota_\epsilon R^a + i \epsilon \gamma^a \psi \ ,
\end{equation}
where we have used $\iota_\epsilon V^a = 0 = \iota_\epsilon \omega^{ab}$ since they have no ``legs" in the fermionic directions. We can repeat the same manipulations for $\omega$ and $\psi$ in order to obtain
\begin{equation}\label{ETM}
	\tilde{\delta}_\epsilon \omega^{ab} = \iota_\epsilon R^{ab} - \frac{1}{\lambda} \bar{\epsilon} \gamma^{ab} \psi \ , \ \tilde{\delta}_\epsilon \psi = \iota_\epsilon \Sigma + \mathcal{D} \epsilon + V^a \gamma_a \epsilon \ .
\end{equation}
We can now compute the difference between the supersymmetry transformations and the superdiffeomorphisms along fermionic directions:
\begin{align}
	\left( \delta - \tilde{ \delta} \right) V^a & = - \iota_\epsilon R^a \ , \\
	\left( \delta - \tilde{ \delta} \right) \omega^{ab} & = - \iota_\epsilon R^{ab} \ , \\
	\left( \delta - \tilde{ \delta} \right) \psi & = - \iota_\epsilon \Sigma \ .
\end{align}
Notice that once these transformations are inserted into the Lagrangian we obtain
\begin{equation}
	\left( \delta - \tilde{\delta} \right) L = \iota_{\epsilon} \left( R^a R^{bc} \epsilon_{a bc } + 2 i \bar{\Sigma} \Sigma \right) = \iota_\epsilon d L \ ,
\end{equation}
which is the contraction of the exterior derivative of the Lagrangian discussed in section 3.3. As mentioned in section 4.4, these trivial gauge transformations vanish identically if the Lagrangian is closed, which is the condition that we discuss in the following section while constructing super Chern-Simons theory on supergroups.

\sect{Supersymmetric Achucarro-Townsend Model}

The model discussed up to this point shows a gauge supersymmetry which is translated into a local 
supersymmetry in the supergravity re-interpretation. Now, we would like to choose a different 
path and before rewriting the Achucarro-Townsend Chern-Simons theory in terms of vielbein $V^a$, the spin connection $\omega^{ab}$ and 
the gravitino $\psi$, we promote it to a worldvolume supersymmetric Chern-Simons model. 
For that, we introduce the worldvolume vielbein $e^a$ and a worldvolume gravitino $\chi^{\alpha}$ (where the index $\alpha=1,2$ denotes the two components of the wordvolume spinor) such that 
\begin{eqnarray}
\label{SAT}
d e^a = \bar{\chi} \gamma^a \chi\,, ~~~~~ d \chi^{\alpha} = 0\,, 
\end{eqnarray}
i.e. assume a flat worldvolume. To translate the CS action given in \eqref{action1} into a worldvolume supersymmetric model, we recall the properties of the $OSp(1|2)$ super algebra (we consider here only the $OSp$ part of the CS action \eqref{action1}); it can be described in terms of 
the generators $T_a, Q_\alpha$ with the commutators 
\begin{eqnarray}
\label{SLA}
[T_a, T_b] =i \epsilon _{ab}^{~~~c} T_c\,,~~~~~~~
[T_a, Q_\alpha] = \gamma_{a\alpha}^\beta Q_\beta\,, ~~~~~~
\{Q_\alpha, Q_\beta\} = 2 i \gamma^{a}_{\alpha\beta} T_a\,. 
\end{eqnarray}
The indices are $a = 1, \dots, 3$ and $\alpha, \beta=1,\dots, 2$. 
The invariant tensors are defined as follows 
\begin{eqnarray}
\label{SLB}
{\rm Str}(T_a T_b) = \eta_{ab}\,, ~~~~~~
{\rm Str}(Q_\alpha Q_\beta) = \epsilon_{\alpha\beta}\,, ~~~~~~
{\rm Str}(T_a Q_\alpha) =0\,. 
\end{eqnarray}
The first one is symmetric $\eta_{ab} = \eta_{ba}$ and the second one is anti-symmetric
$\epsilon_{\alpha\beta} = - \epsilon_{\beta\alpha}$. They are both invertible.
%\begin{eqnarray}
%\label{SLC}
%k_{AB} \tau^A_{(IJ} \tau^B_{K)L} =0\,, 
%~~~~~
%\tau^B_{IJ} f_{BA}^{~~~C} + k_{AB} 
%(
%\tau^{B}_{JK} \omega^{KL} \tau^C_{LI} + 
%\tau^{B}_{IK} \omega^{KL} \tau^C_{LJ}
%) =0
%\,. 
%\end{eqnarray}
%are satisfied. 

The supergroup connection $\bf \Omega$ is defined as \eqref{Omdef}
\begin{eqnarray}
\label{SLD}
{\bf \Omega} = A^a T_a + \psi^\alpha Q_\alpha\,. 
\end{eqnarray}
To respect the statistics of ${\bf \Omega}$, $A^a$ is an anticommuting 
connection (bosonic 1-form) and $\psi^\alpha$ is a commuting connection (fermionic 1-form). 
Promoting both $A^a$ and $\psi^\alpha$ to $(1|0)$-superforms (for a discussion on forms on supermanifolds see e.g. \cite{if1}), they read
\begin{eqnarray}
\label{SLE}
A^a = A^a_b e^b + A^a_\beta \chi^\beta\,, ~~~~~
\psi^\alpha = \psi^\alpha_b e^b + \psi^\alpha_\beta \chi^\beta\,, 
\end{eqnarray}
where $A^a_b$ and $\psi^\alpha_\beta$ are commuting superfields and 
$A^a_\alpha$ and $\psi^\alpha_b$ anticommuting superfields. 
The field strengths are defined as (for simplicity we set $\lambda = 1$ in the following and to distinguish them from the $x$-dependent $\mathbf{\Omega}$ connection, we will rename the superspace-dependent connection with $\cal A$)
\begin{eqnarray}
\label{SLK}
{\cal F} &=& d {\cal A} - [{\cal A}, {\cal A}\}\,, ~~~~~~\nonumber \\
F^a &=& d A^a + \epsilon_{~bc}^{a} A^b \wedge A^c - \frac12  
\psibar \gamma^a \psi\,, ~~~~\nonumber \\
F^\alpha &=& d \psi^\alpha - A^a  (\gamma_a \psi)^\alpha\,.
\end{eqnarray}
Notice that \eqref{SLK} coincide with the $OSp(1|2)$ part of \eqref{defR} and \eqref{defSigma} but hold on the whole superspace. Their rheonomic parameterizations are the following
\begin{eqnarray}
\label{SLH}
{\cal F} &=& {\cal F}_{ab} e^a\wedge e^b + ({\cal W} \gamma_a \chi) \wedge e^a\,, \nonumber \\
\nabla {\cal W}_\alpha &=& e^a \nabla_a {\cal W}_\alpha - \frac14 (\gamma^{ab} \chi)  {\cal F}_{ab}\,, \nonumber \\
F^a &=& F^a_{bc} e^b\wedge e^c + (\Xi^a \gamma_b \chi) \wedge e^b\,, \nonumber \\
\nabla \Xi^a &=& e^b \nabla_b \Xi^a - (\frac14 \gamma^{bc}\chi) F^a_{bc}\,, \nonumber \\
F^\alpha  & =& F^\alpha_{ab} e^a\wedge e^b + (G^\alpha \gamma_a \chi) \wedge e^a\,, \nonumber \\
\nabla G^\alpha &=& e^a \nabla_a G^\alpha - (\frac14 \gamma^{ab}\chi) F^\alpha_{ab}\,,  
\end{eqnarray}
and satisfy the Bianchi identities 
\begin{eqnarray}
\label{SLHA}
d {\cal F} + [{\cal A}, {\cal F}\}=0\,.
\end{eqnarray}
The superfield ${\cal W}$ is a 0-form spinor superfield and its components are defined as
\begin{eqnarray}
\label{SLFA}
{\cal W} = \Xi^a T_a + G^\alpha Q_\alpha\,. 
\end{eqnarray}
The fields $\Xi^a$ (anticommuting) and $G^\alpha$ (commuting) are 0-form woldvolume spinors 
with indices in the vector and spinor representation of $SO(1,2)$, respectively.
Their covariant derivatives are defined as follows
\begin{eqnarray}
\label{SLHB}
\nabla {\cal W} &=& d {\cal W} + [{\cal A}, {\cal W}\} \nonumber \\
(\nabla \Xi)^a &=& d \Xi^a + \epsilon_{~bc}^{a} A^b \Xi^c + \gamma^a_{\alpha\beta} G^\alpha \psi^\beta\,, \nonumber \\
(\nabla G)^\alpha &=& d G^\alpha - \gamma^\alpha_{~a\beta} (A^a G^\beta + \Xi^a \psi^\beta) \,. 
\end{eqnarray}

The supersymmetry transformations can be easily obtained by using the rheonomic parameterization (\ref{SLH})
\begin{eqnarray}
\label{SLI}
\delta_\epsilon A^a &=& {\cal L}_\epsilon A^a = 
\iota_\epsilon F^a +  \nabla \Lambda^a = 
(\Xi^a \gamma_b \epsilon) e^b + \nabla \Lambda^a\,, ~~~~\nonumber \\
\delta_\epsilon \psi^\alpha &=& {\cal L}_\epsilon \psi^\alpha =   
\iota_\epsilon F^\alpha  + \nabla \Lambda^\alpha = (G^\alpha \gamma_a \epsilon) e^a + \nabla \Lambda^\alpha\,,
\end{eqnarray}
where $\Lambda^a = \iota_\epsilon A^a$ and $\Lambda^\alpha = \iota_\epsilon \psi^\alpha$. 
The fields $\Xi^a$ and $G^\alpha$ are the superpartners of $A^a$ and $\psi^\alpha$. No auxiliary field is needed and the matching is achieved both off-shell and on-shell. All the symmetries close off shell because gauge symmetries close by construction in SCS theories and the supersymmetries close off-shell in $d=3$ supergravity.

The action invariant under both gauge and worldvolume symmetries is given by \cite{GrassiMaccaferri}
\begin{eqnarray}
\label{SLF}
{\cal L}^{(3|0)} =  {\rm Str} \left( {\cal A} d {\cal A} - \frac23 {\cal A} [{\cal A}, {\cal A}\} + 
{\cal W}_\alpha {\cal W}_\beta \epsilon^{\alpha\beta} \text{Vol}^3 \right)
\end{eqnarray}
$\text{Vol}^3$ is the volume form on the three-dimensional manifold,
$\text{Vol}^3 = \frac{1}{3!} \epsilon_{abc} e^a e^b e^c.$\footnote{$d \text{Vol}^3 = \frac12 \epsilon_{abc} 
\bar\chi \gamma^a \chi e^b e^c = \frac12 \rho_\alpha \epsilon^{\alpha\beta} \rho_\beta$ with 
$\rho_\alpha = (V^a \gamma_a \psi)_\alpha$ and 
$d \rho_\alpha =0$.} We use the notation 
${\cal W} \cdot {\cal W} = {\cal W}^\alpha \epsilon_{\alpha\beta} {\cal W}^\beta$

Computing the super trace explicitly we have 
\begin{eqnarray}
\label{SLG}
{\cal L}^{(3|0)} &=&\left( 
\eta_{ab} A^a d A^b - \frac23 \epsilon_{abc} A^a A^b A^c + 
\eta_{ab} \Xi^a \cdot \Xi^b \text{Vol}^3 \right. \nonumber \\
&+& \left. \epsilon_{\alpha \beta} \psibar^\alpha d \psi^\beta 
+  A^a \bar\psi \gamma_a \psi + 
\epsilon_{\alpha\beta} G^\alpha \cdot G^\beta \text{Vol}^3 \right)\,. 
\end{eqnarray}

The Chern-Simons gauge symmetries are given by
\begin{eqnarray}
\label{SLJ}
\delta {\cal A} = d C + [{\cal A}, C\}\,, ~~~~~~ 
\delta {\cal W} = [{\cal W}, C\}\,, ~~~~~
%s C = \frac12 [C, C\}\,, ~~~~~
C = C^a T_a + C^\alpha Q_\alpha\,. 
\end{eqnarray}
where $C$ is the gauge superfield parameter with values in the super-Lie algebra.

\begin{comment}
The nilpotency of the BRST 
operator implies that 
\begin{eqnarray}
\label{SLJ_A}
s C &=&\frac12  [C, C\}\,, \nonumber \\
s C^a &=& \frac12 \epsilon^{a}_{~bc} C^b C^c + \frac12 \gamma^a_{\alpha \beta} C^\alpha C^\beta\,, ~~~~~ 
\nonumber \\
s C^\alpha &=& \gamma^\alpha_{a\beta} C^a C^\beta\,.   
\end{eqnarray}
which is nilpotent because of the Jacobi identities. 
\end{comment}

Let us compute the differential of ${\cal L}^{(3|0)}$
\begin{eqnarray}
\label{SL_K}
d {\cal L}^{(3|0)} &=& {\rm Str}\left( {\cal F} \wedge {\cal F} + 
2 d {\cal W} \cdot {\cal W} \text{Vol}^3 + \frac12 {\cal W}^2 \epsilon_{abc} 
(\bar\chi\gamma^a \chi) e^b e^c\right) \nonumber \\
&=& {\rm Str}\left( 
2  {\cal F}_{ab} e^a\wedge e^b  ({\cal W} \gamma_c \chi) \wedge e^c + 
({\cal W} \gamma_a \chi) \wedge e^a ({\cal W} \gamma_b \chi) \wedge e^b \right.  \nonumber \\
&+& \left. 2 
(- \frac14 (  {\cal W} \gamma^{ab} \chi)  \text{Vol}^3 {\cal F}_{ab}+ \frac12 {\cal W}^2 \epsilon_{abc} 
(\bar\chi\gamma^a \chi) e^b e^c\right)  =0
\end{eqnarray}
which agrees with the fact that the rheonomic Lagrangian is closed if there are auxiliary fields or 
if they are not needed. We get also the interesting equation 
\begin{eqnarray}
\label{SL_H}
{\rm Str} \left(  {\cal F} \wedge {\cal F}\right) = - d \left({\rm Str} ({\cal W}^2) \text{Vol}^3 \right) 
\end{eqnarray}
Both members are gauge invariant under super gauge transformations. 
Since we have $ d {\cal L}^{(3|0)}=0$, this implies 
\begin{eqnarray}
\label{SL_I}
\delta d {\cal L}^{(3|0)}=0 \Longrightarrow \delta {\cal L}^{(3|0)} =  d \Omega^{(2|0)}_1
\end{eqnarray}
where $\Omega^{(2|0)}_1$ is a $(2|0)$ form. In turn, this 
implies that the action is gauge invariant up to boundary terms. 

We can now add the other part of the supergroup: we consider as in the previous sections $OSp(1|2) \times Sp(2)$ and this rerequires to subtract a super Chern-Simons sction (i.e. with worldvolume supersymmetry) to the action \eqref{SLG}:
\begin{eqnarray}\label{SL_J}
	{\cal L}^{(3|0)} &=&\left( 
\eta_{ab} A^a d A^b + \frac23 \epsilon_{abc} A^a A^b A^c + 
\eta_{ab} \Xi^a \cdot \Xi^b \text{Vol}^3 \right. \nonumber \\
\nonumber &+& \left. \epsilon_{\alpha \beta} \psibar^\alpha d \psi^\beta 
+  A^a \bar\psi \gamma_a \psi + 
\epsilon_{\alpha\beta} G^\alpha \cdot G^\beta \text{Vol}^3 \right) + \\
&-& \left( \eta_{ab} \tilde{A}^a d \tilde{A}^b + \frac{2}{3} \epsilon_{abc} \tilde{A}^a \tilde{A}^b \tilde{A}^c + 
\eta_{ab} \tilde{\Xi}^a \cdot \tilde{\Xi}^b \text{Vol}^3 \right) \ .
\end{eqnarray}
Again the Lagrangian is closed because of the presence of the "auxiliary fields" $\Xi, G \text{ and } \tilde{\Xi}$ and its exterior derivative takes the form (using the parametrisations \eqref{SLH} when calculating $d \Xi$)
\begin{align}\label{SL_KA}
	0 = d {\cal L}^{(3|0)} =& F \wedge F - \tilde{F} \wedge \tilde{ F} + \left( \nabla \psi \right)^2 -\frac12 \text{Vol}^3 \chi \gamma^{cd} \eta_{ab} \left( \Xi^a F^{b}_{cd} - \tilde{\Xi}^a \tilde{F}^{b}_{cd} \right) + \\
	\nonumber &+ \eta_{ab} \left( \Xi^a \Xi^b - \tilde{\Xi}^a \tilde{\Xi}^b \right) 3 \epsilon_{a b c} \bar{\chi} \gamma^a \chi e^b e^c - \frac12 \epsilon_{\alpha \beta} \gamma^{ab} \chi F^\alpha_{ab} G^\beta \text{Vol}^3 + \\
	\nonumber &+ \epsilon_{\alpha \beta} G^\alpha G^\beta 3 \epsilon_{a b c} \bar{\chi} \gamma^a \chi e^b e^c \ .
\end{align}
If we set $\Xi = \tilde{\Xi}$, we have that the first term of the second line vanishes, while the last term of the first line becomes
\begin{equation}\label{SL_L}
	\frac12 \text{Vol}^3 \chi \gamma^{cd} \eta_{ab} \Xi^a \left( F^{b}_{cd} - \tilde{F}^{b}_{cd} \right) \ .
\end{equation}
This term vanishes if
\begin{equation}\label{SL_M}
	F_{ab} = \tilde{F}_{ab} + \bar{\chi} \gamma_{ab} \chi \ ,
\end{equation}
because of the Fierz identities \eqref{Fierz3d}. Once we identify the difference of the gauge fields with the dreibein $V$ as in \eqref{Acombinations}, \eqref{SL_M} is exactly the vanishing torsion-condition, as we can see from \eqref{TORSION}:
\begin{equation}
	F_{ab} = \tilde{F}_{ab} + \bar{\chi} \gamma_{ab} \chi \implies R^a = 0 \ .
\end{equation}
Consider the counting of the degrees of freedom. For the fields of starting Lagrangian counting is
\begin{align*}
	& A: \ \ 9 - 3 = 6\,, \ \ \ \ \ \ \ \ \Xi: \ \ 6 \\
	& \Psi: \ \ 6 - 2 = 4\,, \ \ \ \ \ \ \  G: \ \ 4 \\
	& {\!\!\!\!\!\tilde{A}: \ \ 9 - 3 = 6\,, \ \ \ \  \ \ \ \ \tilde{\Xi}: \ \ 6}\over{\ \ \ \ \ \ \ \ \ 8 \ \ \ \ \ \ \ \ \ \ \ \ \ \ \ \ \ \ \ \ \ \ \   8}  \ 
\end{align*}
The matching is established by construction since the Lagrangian itself has been built by associating a partner to each field. The condition $\Xi = \tilde{\Xi}$ removes $6$ d.o.f. on the right, but since it implies the torsionless condition it removes $3$ d.o.f. on the left as well. Now, the counting is $5$ vs $2$. The matching is established by requiring that out of the $4$ d.o.f. of the auxiliary field $G$ only $1$ is nontrivial; this is obtained via the parametrisation
\begin{equation}
	G^\alpha_{\alpha'} = G \delta^\alpha_{\alpha'} \ .
\end{equation}
This field $G$ can be identified with auxiliary field $f$ of section 4.

\sect{Supersymmetry}

We clarify some issues regarding supersymmetry matching of d.o.f. 
We also point out that here we do not take a supergravity interpretation of Chern-Simons theory, but 
we explore the matching of d.o.f.'s as in a pure Chern-Simons gauge theory. Later, we discuss its supergravity 
interpretation and discuss different supergroups. 

First consider the supersymmetry on the worlvolume. In that case, for each  gauge field $A^a$ 
there is a corresponding spinor field $\Xi^a$. The matching off-shell is achieved  
by noting that, because of the gauge symmetry $\delta A^a = \nabla C^a$, we can remove one 
degree of freedom (for each generator of the Lie algebra) from $A^a$. Then, the remaining d.o.f.'s matche with those of the gauginos $\Xi^\alpha$. On the other hand,
using the equations of motion, we find that there are no propagating d.o.f. for the 
gauge fields, since their field strength vanishes, and for the gauginos which have algebraic 
(non dynamical) equations. 
 
The same argument applies also in the case of the gauge fields $\psi^\alpha$ associated to the supercharges. 
 Gauge symmetry removes one d.o.f. from $\psi^\alpha$ and the remaining d.o.f.
  match those of $G^\alpha$. 
 Again, on-shell there are no propagating degrees of freedom and the matching is trivial. 
 
We consider now a different type of matching. We would like to compare the d.o.f.'s of the gauge fields 
 $A^a$ with those of $\psi^\alpha$. They are both gauge fields and off-shell correspond $2 \times {\it
 bosonic~generators}$ and $2 \times {\it fermionic~generators}$ of the gauge supergroup. Consider then the 
 supergroup ${OSp}(p|q)$. The counting of bosonic vs fermionic generators gives 
 \begin{eqnarray}
\label{bofe_A}
\left( \frac{p(p-1)}{2} + \frac{q(q+1)}{2} \Big|  p q \right)
\end{eqnarray}
and therefore the matching is achieved when $p =q$ or $p = q+1$. In that case, the bosonic and 
fermionic d.o.f.'s match. For example the case $q=2$ and $p=2$ or $p=3$ are example of $d=3$ supergravities 
already known in the literature. However, in general, the matching is not achieved. 

If $p\neq q$ we have to follow a different path. Assuming that $p>q$, then 
we have that the super coset ${OSp}(p|q)/SO(p-q)$ has the same number 
of bosonic and fermionic generators  $(p q | p q)$. On the other side, 
if $p<q$, we find that the supercoset ${OSp}(p|q)/Sp(q-p)$
has the same number of bosonic and fermionic generators. Lastly, we 
have the case $p=1$ and $q = 2 r$. In that case, the coset with the matching is 
${OSp}(1| 2 r)/SO(r,r)$. 

We can distinguish the bosonic gauge fields $A^a$ between those with the index $I = 1, \dots, pq$ (the number 
of fermionic generators) $A^I$ and those belonging to the subgroups $SO(p-q)$, or $Sp(q-p)$ or 
$SO(r,r)$ (depending on $p>q$, $p<q$ or $p=1, q=2 r $)  $A^i$ where $i$ runs over the generators 
of the subgroup). Then, the field strengths can be divided as follows
\begin{eqnarray}
\label{bofe_B}
F^I = d A^I + f^{I}_{i J} A^i\wedge A^J = \nabla A^I\,, ~~~~~
F^i = d A^i + f^{i}_{IJ} A^I \wedge A^J + f^i_{jk} A^j \wedge A^k\,. 
\end{eqnarray}
The torsion condition
the equation 
\begin{eqnarray}
\label{bofe_C}
F^I = 0\,, 
\end{eqnarray}
can be solved in terms of $A^i$, the gauge fields of the subgroup. 
In this way, by going (partially) on-shell with those degrees of freedom, we achieve the off-shell matching 
for the remaining d.o.f.'s. Once this equation is solved in terms of $A^i$, we can reinstate them 
in the rest of the action and derive the corresponding equations of motion. 
 Through the Bianchi identities, we have that 
\begin{eqnarray}
\label{bofe_D}
\nabla F^I = f^I_{i J} F^i\wedge A^J 
\end{eqnarray}
and therefore, imposing $F^I=0$, we find a condition on $A^I$ (this is analogous to impose 
the vanishing of the torsion in general relativity, solving the spin connection in terms of the vielbein; 
then, this implies the condition $E \wedge R =0$ for the curvature). This corresponds to 
the reduced holonomy 
\begin{eqnarray}
\label{bofe_E}
 f^I_{i J} F^i\wedge A^J =0\,. 
\end{eqnarray}
 Even in this reduced holonomy situation, eq.(\ref{bofe_C}) is not always solvable. 
 Indeed, counting the independent contained in (\ref{bofe_C}) we have that the index $I$ runs from 1 to $pq$, but they 
 are 2-form equations which have $pq (pq -1)/2$ independent components. The unknowns given 
 by the gauge fields $A^i$ are 1-form (with $pq$ components for each value of the index $i$), therefore 
 there are $pq (p-q)^2 (p-q-1)/2$ unknowns. To solve the equations we need the matching 
 \begin{eqnarray}
\label{matA}
{p^2 q^2 (pq -1)} = pq (p-q) (p-q-1)
\end{eqnarray}
which can be achieved only if 
$p=1, \forall q$, or $q=-1, \forall p$ and $q = p/(p+1), \forall p\neq -1$. 
However, $p$ and $q$ must be positive 
integers, this excludes the last two solutions. The remaining one, $p=1$, is the only possible case for any 
$q$. This corresponds to the case $OSp(1|2 r)$ with the subgroup $SO(r,r)$ which is a subgroup of 
$Sp(2r)$.  In that case we can solve the equation (\ref{bofe_C}) in terms of $A^i$. 
We have the interesting cases $OSp(1|2)/SO(1,1), OSp(1|4)/SO(2,2), OSp(1|6)/SO(3,3)$ and 
the $OSp(1|32)/SO(16,16)$. 

\section{Conclusions and outlook}

In this note we have clarified the issues regarding the relation between the Achucarro-Tonwsend supergravity models and the group manifold approach to the same theory. In the 
first case, the gauge symmetry is promoted to a super gauge symmetry and therefore closes off-shell. In the 
second case, the supersymmetry closes off-shell only after the introduction of auxiliary fields. In the second part of the paper, 
we construct a double supersymmetric version with worldvolume and gauge supersymmetry and discuss how supergravity can be retrieved. The present work prepares the way to construct Achucarro-Tonwsend supergravities 
with extended supersymmetries corresponding to orthosymplectic groups $OSp(p|2) \times OSp(q|2)$. 
Only few cases are studied (see for example \cite{tartaglino-kuzenko,Concha}) in the superspace language, but 
not in the group manifold approach. In the latter, the question of auxiliary fields has never been tackled. 

\sect{Appendix A : gamma matrices in $D=3$}

We summarize in this Appendix our gamma matrix conventions in $D=3$.

\eq
 \ga_0 =
\left(
\begin{array}{cc}
  i &  0    \\
  0&  -i
\end{array}
\right),~~~\ga_1=
\left(
\begin{array}{cc}
  0&   1  \\
  1&     0
\end{array}
\right)
,~~~\ga_2=
\left(
\begin{array}{cc}
  0&   -i \\
  i&     0
\end{array}
\right)
\en

\eqa
& & \eta_{ab} =(-1,1,1),~~~\{\ga_a,\ga_b\}=2 \eta_{ab},~~~[\ga_a,\ga_b]=2 \ga_{ab}= -2 \epsi_{abc} \ga^c, \\
& & \epsi_{012} = - \epsi^{012}=1, \\
& & \ga_a^\dagger = \ga_0 \ga_a \ga_0,~~\ga_a^T= - C \ga_a C^{-1}, ~~C^T = -C, ~~C^2 = \onebold
\ena

\subsection{Useful identities}

\eqa
 & &\ga_a\ga_b= \ga_{ab}+\eta_{ab}= - \epsi_{abc} \ga^c + \eta_{ab}\\
 & &\ga_{ab} \ga_c=\eta_{bc} \ga_a - \eta_{ac} \ga_b -\epsi_{abc}\\
 & &\ga_c \ga_{ab} = \eta_{ac} \ga_b - \eta_{bc} \ga_a -\epsi_{abc}\\
 & &\ga_a\ga_b\ga_c= \eta_{ab}\ga_c + \eta_{bc} \ga_a - \eta_{ac} \ga_b - \epsi_{abc}\\
 & &\ga^{ab} \ga_{cd} = - 4 \de^{[a}_{[c} \ga^{b]}_{~~d]} - 2 \de^{ab}_{cd}
  \ena
 \noi where $\de^{ab}_{cd}
 = \unmezzo (\de^a_c \de^b_d - \de^a_d \de^b_c)$, and index antisymmetrizations in square brackets have weight 1.

\subsection{Fierz identity for two Majorana one-forms}

\eq
\psi \psibar = {1 \over 2} (\psibar \ga^a \psi ) \ga_a
\en
As a consequence 
\eq
\ga_a \psi \psibar \ga^a \psi =0   \label{Fierz3d}
\en

\section*{Acknowledgement}

This work has been partially supported by Universit\`a del Piemonte Orientale research funds. We would like to thank L. Andrianopoli and M. Trigiante for useful discussions on supergravity.

\vfill\eject
\end{document}